\documentclass{article}
\usepackage{spconf,amsmath,graphicx}

\usepackage{amsmath,graphicx}
\usepackage{pifont}
\newcommand{\cmark}{\ding{51}}%
\newcommand{\xmark}{\ding{53}}%
\usepackage{amssymb}
\usepackage{bm}
\usepackage{multirow}
\usepackage{adjustbox}
\usepackage{tabularx}
\usepackage{tikz}
\usepackage{caption}
\usepackage{subcaption}
\usepackage{verbatim}
\usepackage{float}
\usepackage{commath}
\usepackage{url}
\title{Dual-path Self-Attention RNN for Real-Time Speech Enhancement}
\name{Ashutosh Pandey$^\textup{1}$ and DeLiang Wang$^\textup{1,2}$\thanks{This research was supported in part by two NIDCD grants (R01DC012048 and R02DC015521) and the Ohio Supercomputer Center.}}
\address{$^\textup{1}$Department of Computer Science and Engineering, The Ohio State University, USA\\
$^\textup{2}$Center for Cognitive and Brain Sciences, The Ohio State University, USA\\
\texttt{\{pandey.99, wang.77\}@osu.edu}}
\begin{document}
%
\maketitle
\begin{abstract}
We propose a dual-path self-attention recurrent neural network (DP-SARNN) for time-domain speech enhancement. We improve dual-path RNN (DP-RNN) by augmenting inter-chunk and intra-chunk RNN with a recently proposed efficient attention mechanism. The combination of inter-chunk and intra-chunk attention improves the attention mechanism for long sequences of speech frames. DP-SARNN outperforms a baseline DP-RNN by using a frame shift four times larger than in DP-RNN, which leads to a substantially reduced computation time per utterance. As a result, we develop a real-time DP-SARNN by using long short-term memory (LSTM) RNN and causal attention in inter-chunk SARNN. DP-SARNN significantly outperforms existing approaches to speech enhancement, and on average takes 7.9 ms CPU time to process a signal chunk of 32 ms. 
\end{abstract}
\section{Introduction}
\label{sec:intro}
Speech enhancement aims at improving the intelligibility and quality of a speech signal corrupted by additive noise. It is used as a preprocessor in many applications, such
as automatic speech recognition, telecommunication, and hearing aids, to improve their performance in noisy environments.

Traditional speech enhancement approaches include spectral subtraction, Wiener filtering and statistical model-based methods \cite{loizou2013speech}. In recent years, speech enhancement has been extensively studied as a deep learning problem \cite{wang2017supervised}. In particular, time-domain speech enhancement is becoming increasingly popular due to its capability to jointly enhance both the magnitude and the phase of noisy speech. Further, time domain approaches do not require transformations to and from the frequency domain. 

Representative time-domain networks include UNet \cite{ronneberger2015u} convolutional neural networks (CNNs) \cite{pandey2019new, pascual2017segan}, CNNs with temporal and dilated convolutions \cite{pandey2019tcnn, rethage2017wavenet}, dense CNN \cite{pandey2020densely}, and CNN with self-attention \cite{pandey2020dense}.

Dual-path recurrent neural network (DP-RNN) was recently proposed for time-domain speaker separation with state-of-the art performance \cite{luo2020dual}. In DP-RNN, a sequence of frames is divided into overlapping chunks and processed by a series of intra-chunk and inter-chunk RNNs. The efficacy of DP-RNN can be attributed to a reduced sequence length per RNN for efficient training, and a smaller frame size (0.25 ms) for speech processing.

Motivated by the success of DP-RNN for speaker separation \cite{luo2020dual} and the attention mechanism for speech enhancement \cite{pandey2020dense}, in this work, we propose to augment DP-RNN with attention. We replace inter-chunk and intra-chunk RNN in DP-RNN with a recently proposed self-attention RNN (SARNN) \cite{merity2019single}. SARNN was proposed as an efficient technique to augment RNNs with attention, resulting in reduced memory consumption, faster training, and state-of-the-art performance in natural language processing. A similar idea of augmenting DP-RNN with attention using a different approach was recently proposed in \cite{tan2020sagrnn} for binaural speaker separation.

Dual-path SARNN (DP-SARNN) outperforms DP-RNN by using a frame shift four times larger than in DP-RNN, resulting in considerably reduced computation time per utterance. We use low-latency DP-SARNN to develop a real-time algorithm by using long short-term memory (LSTM) RNN and causal attention in inter-chunk SARNN.

DP-SARNN significantly outperforms existing methods for speech enhancement for both offline and real-time speech enhancement. Real-time DP-SARNN, on average, takes 7.9 ms CPU time to process a signal chunk of 32 ms, which is significantly less than than other comparable time-domain models, DP-RNN (17.4 ms) and dense convolutional network (DCN) (11.0 ms) \cite{pandey2020dense}.

The rest of the paper is organized as follows. Section 2 describes DP-SARNN. Experimental settings and results are given in Section 3. Section 4 concludes the paper.
\vspace{-7pt}
\section{Model Description}
\subsection{Self-Attention RNN}
SARNN was recently proposed in \cite{merity2019single} for efficiently combining attention mechanism \cite{vaswani2017attention} with recurrent processing of RNN. A block diagram of SARNN used in this study is shown in Fig. 1. 
\begin{figure}[!h]
\centering
\includegraphics[width=\columnwidth, keepaspectratio]{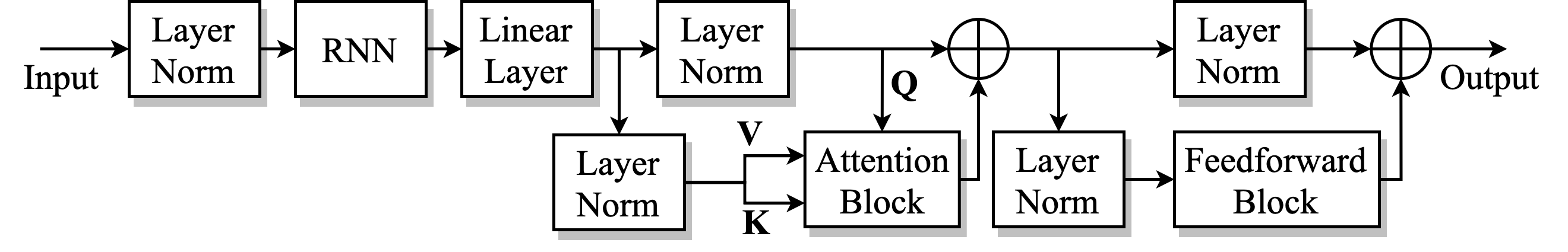}
\caption{The proposed architecture of SARNN.}
\label{fig:rnnt}
\end{figure}

SARNN comprises an RNN, an attention block, and a feedforward block . The input to SARNN is a matrix  $\bm{X} \in \mathbb{R}^{T \times N}$, where $T$ is the sequence length and $N$ is the feature dimension. $\bm{X}$ is layer normalized \cite{ba2016layer} and fed to an RNN of hidden size $H$, which is followed by a linear layer to  transform the RNN output to the same shape as the input. Next, two different layer normalizations are used to get query ($\bm{Q}$), key ($\bm{K}$) and value ($\bm{V}$) for the  following attention block, where $\bm{K}$ is equal to $\bm{V}$. $\bm{Q}$ is added to the output of the attention block to form a residual connection. The output after the attention block is processed using the feedforward block in a residual way as shown in Fig. 1.

A block diagram of the attention block in SARNN is shown in Fig. 2. It comprises three trainable vectors $\{\bm{Q}^{\prime}, \bm{K}^{\prime}, \bm{V}^{\prime}\} \in \mathbb{R}^{1 \times N}$, and its inputs are $\{\bm{Q}, \bm{K}, \bm{V}\} \in \mathbb{R}^{T \times N}$. $\bm{Q}, \bm{K}$, and $\bm{V}$ are refined using a gating mechanism given in the following equation. 
\begin{figure}[!b]
\centering
\includegraphics[width=0.58\columnwidth, keepaspectratio]{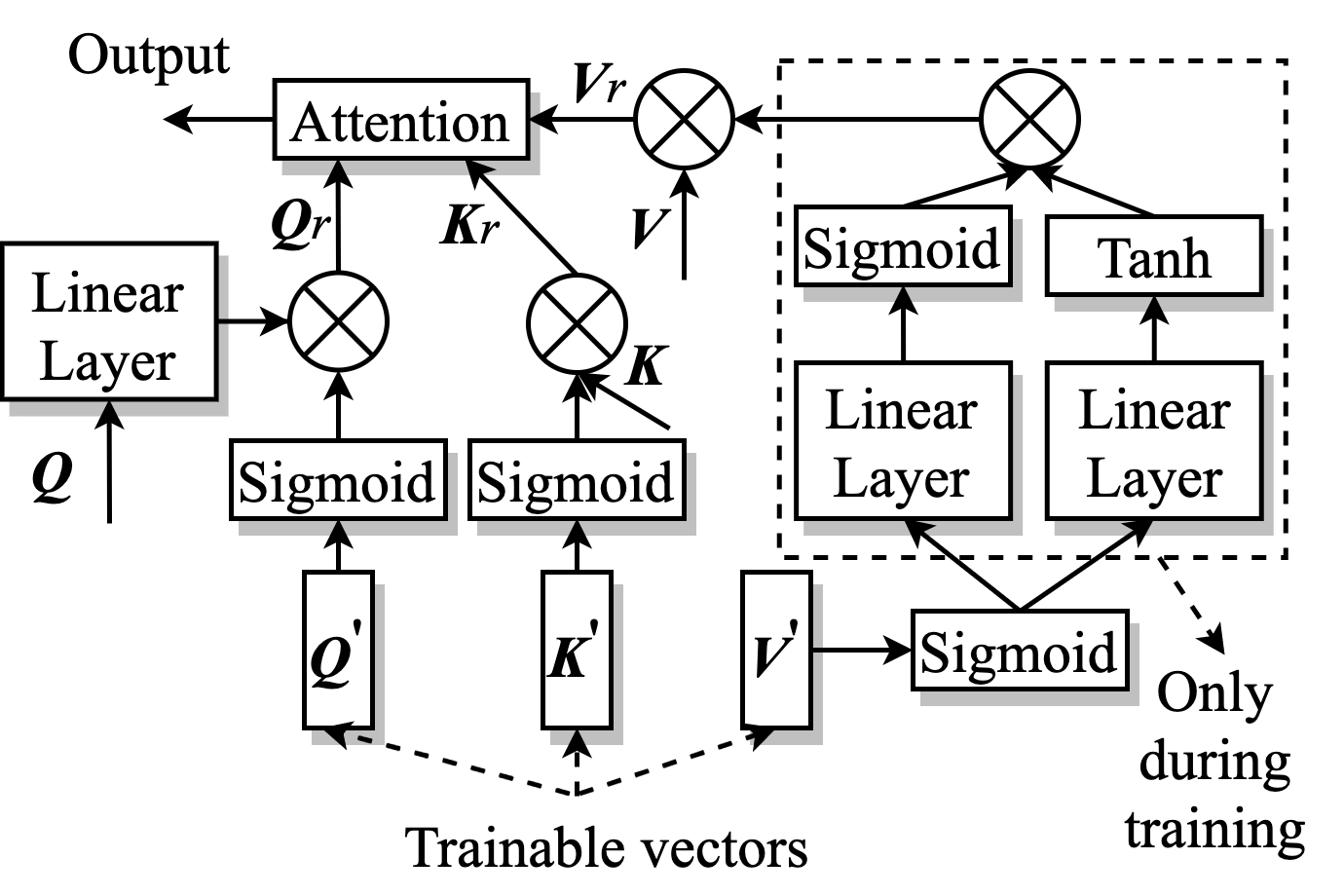}
\caption{Attention block in SARNN.}
\label{fig:rnnt}
\end{figure}
\begin{equation}
\begin{split}
\bm{K}_{r} &= \bm{K} \odot \text{Sigm}(\bm{K}^{\prime})\\
\bm{Q}_{r} &= \text{Lin}(\bm{Q}) \odot  \text{Sigm}(\bm{Q}^{\prime}) \\
\bm{V}_{r} &= \bm{V}\odot  [\text{Sigm}(\text{Lin}(\bm{V}^{\prime})) \odot \text{Tanh}(\text{Lin}(\bm{V}^{\prime}))]
\end{split}
\end{equation}  
where $\text{Sigm()}$ is sigmoidal nonlinearity, $\text{Lin()}$ is a linear layer, and $\odot$ denotes elementwise multiplication. $\bm{Q}^{\prime}, \bm{K}^{\prime}$, and $\bm{V}^{\prime}$ are broadcasted to match the shape of $\bm{Q}, \bm{K}$, and $\bm{V}$. Note that $\text{Sigm}(\text{Lin}(\bm{V}^{\prime})) \odot \text{Tanh}(\text{Lin}(\bm{V}^{\prime}))$ is a deterministic vector, and hence this operation is used only during training to better optimize  $\bm{V}^{\prime}$, and its final value is stored as a vector to use during evaluation.
 
 The final output of the attention block is computed as
 \begin{equation}
 \bm{A} = \text{Softmax}(\frac{\bm{Q}_{r} \bm{K}_{r}^{T}}{\sqrt{N}})\bm{V}_{r}
 \end{equation}
 \vspace{-10pt}
The feedforward block in SARNN, shown in Fig. 3, is similar to the ones used in transformer networks \cite{vaswani2017attention}. It is a fully connected network with one hidden layer of size 4N with Gaussian error linear unit (GELU) nonlinearity \cite{hendrycks2016gaussian} and dropout. 

\begin{figure}[!t]
\centering
\includegraphics[width=0.7\columnwidth, keepaspectratio]{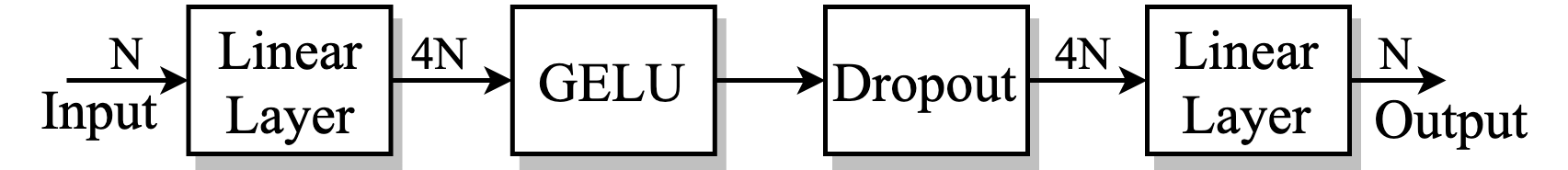}
\caption{Feedforward block in SARNN.}
\label{fig:rnnt}
\end{figure}
\subsection{Dual-Path SARNN}
Let $\bm{X} \in \mathbb{R}^{T \times N}$ be a matrix representing a time series signal $\{\bm{x}_{1} \cdots \bm{x}_{T}\}$, where $\bm{x}_{i} \in \mathbb{R}^{N \times 1}$. We can divide $\bm{X}$ into overlapping chunks of length $K$  with a chunk shift of $P$ and concatenate them together to get a 3D tensor $\mathbf{X} \in \mathbb{R}^{J \times K\times N}$, where $J$ is the number of chunks. If required, $\bm{X}$ is zero padded to the right to get the last chunk of size $K$. The $k^{th}$ signal in the $j^{th}$ chunk of $\mathbf{X}$ is defined as
\begin{equation}
\mathbf{X}_{j, k} = \bm{x}_{(j-1)*P + k}, \ 1 \leq j \leq J, 1 \leq k \leq K
\end{equation}
DP-SARNN is modeled by replacing RNNs in a DP-RNN \cite{luo2020dual} with SARNNs. An illustrative diagram of DP-SARNN is shown in Fig. 4. A DP-SARNN comprises two SARNNs: intra-chunk SARNN and inter-chunk SARNN. It takes $\mathbf{X}$ as input and processes it using intra-chunk SARNN and inter-chunk SARNN in that order. Intra-chunk SARNN considers frames in a chunk as the sequential input, and separately processes all the chunks $[\mathbf{X}_{1} \cdots \mathbf{X}_{J}]$ and concatenates them together to get $\mathbf{X}^{1} \in \mathbb{R}^{J \times K \times N}$. Next, $\mathbf{X}^{1}$ is transposed along the first and second dimension to get $\mathbf{X}^2 \in  \mathbb{R}^{K \times J \times N}$. $\mathbf{X}^{2}$ is fed to inter-chunk SARNN, which considers chunks as the sequential input, and separately processes $[\mathbf{X}^{2}_{1} \cdots \mathbf{X}^{2}_{K}]$ and concatenates them together to get $\mathbf{X}^{3} \in \mathbb{R}^{K \times J \times N}$. Finally, $\mathbf{X}^{3}$ is transposed to get the final output $\mathbf{X}^{4} \in \mathbb{R}^{J \times K \times N}$. 
\begin{figure}[!h]
\centering
\includegraphics[width=0.98\columnwidth, keepaspectratio]{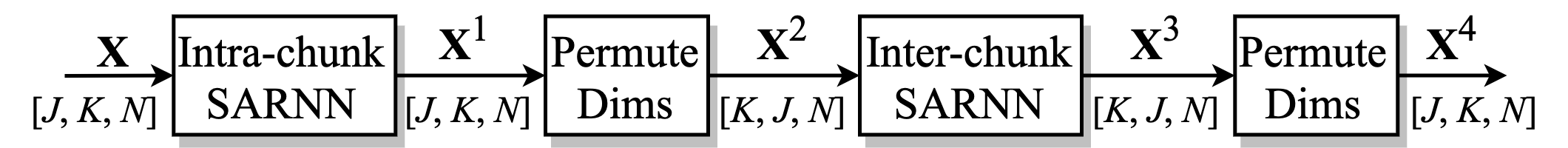}
\caption{An illustrative diagram of DP-SARNN.}
\label{fig:rnnt}
\end{figure}
\begin{figure*}[!t]
\centering
\includegraphics[width=0.82\textwidth, keepaspectratio]{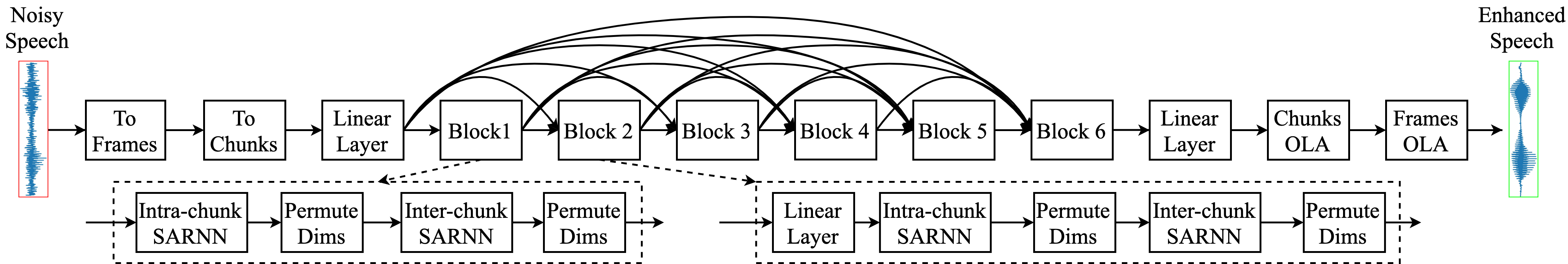}
\caption{The proposed speech enhancement network.}
\label{fig:rnnt}
\end{figure*}
\vspace{-10pt}
\subsection{Speech enhancement using DP-SARNN}
Given a speech signal $\bm{s}$, a noise signal $\bm{n}$, the noisy speech signal is modeled as
\begin{equation}
\bm{x} = \bm{s} + \bm{n}
\end{equation}
where $\{\bm{x}, \bm{s}, \bm{n}\} \in \mathbb{R}^{M \times 1}$, and $M$ represents the number
of samples in the signal. The goal of a speech enhancement algorithm is to get a close estimate, $\bm{\hat{s}}$, of $\bm{s}$ given $\bm{x}$.

$\bm{x}$ is first converted to frames using a frame size of $L$ and frame shift of $R$ to get $\bm{X} \in \mathbb{R}^{T \times L}$, where $T$ is the number of frames.  Next, $\bm{X}$ is divided into chunks with a chunk size of $K$ and chunk shift of $P$ to get $\mathbf{X} \in \mathbb{R}^{J \times K \times L}$, where $J$ is the number of chunks.

$\mathbf{X}$ is fed to a DP-SARNN based speech enhancement network shown in Fig. 4. The proposed network consists of a linear layer at the input, six DP-SARNN blocks, and one linear layer at the output. The input linear layer is used to project $\mathbf{X}$ to a higher dimension of size $N$. It is then processed by six DP-SARNN blocks. The input to a given DP-SARNN block is a concatenation of the outputs of all the previous blocks and the input layer. If the input size to a block is greater than $N$, it is projected to size $N$ using a linear layer. Finally, the linear layer at the output transforms the output dimension from $N$ to $L$ (frame size) to get $\mathbf{\widehat{S}} \in \mathbb{R}^{J \times K \times L}$. $\mathbf{\widehat{S}}$ is subjected to overlap-and-add (OLA) along the chunks to get  $\bm{\widehat{S}} \in \mathbb{R}^{T \times L}$. $\bm{\widehat{S}}$ is subjected to OLA along the frames to get enhanced signal $\bm{\hat{s}} \in \mathbb{R}^{M \times 1}$.
\subsubsection{Real-time speech enhancement using DP-SARNN} 
A real-time speech enhancement system must satisfy a causality constraint and a latency constraint. The causality constraint requires that the output for a given frame is computed using only the current and the previous frames. For DP-SARNN, we modify the frame-level constraint to chunk-level as given in Eq. 5.
\begin{equation}
\mathbf{\widehat{S}}_{j} = f_{\theta}([\mathbf{X}_{1}, \mathbf{X}_{2} \cdots \mathbf{X}_{j-1}, \mathbf{X}_{j}])
\end{equation} 
Where $f_{\theta}$ is a DP-SARNN model parametrized by $\theta$.

The latency constraint requires the frame size to be small and the computation time for a frame to be less than the frame shift \cite{reddy2020icassp}. Similar to causality constraint, we apply latency constraint to chunks for DP-SARNN. 

For non-causal speech enhancement, we use bidirectional long short-term memory (BLSTM) RNN in inter-chunk and intra-chunk SARNN. For causal speech enhancement, the BLSTM in inter-chunk SARNN is replaced with LSTM, and the attention defined in Eq. (2) is replaced with a causal attention defined in Eq. (7).
\begin{equation}
\begin{split}
\bm{W} &=\text{Softmax}(\frac{\bm{Q}_{r} \bm{K}_{r}^{T}}{\sqrt{N}})\\
\bm{A}_{causal} &= \text{Mask}(\bm{W})\bm{V}_{r}
\end{split}
\end{equation}
where $\bm{W} \in \mathbb{R}^{T \times T}$ , and 
\begin{equation}
	\text{Mask}(W)(i, j) = \begin{cases}
	      W(i, j), & \text{if}\ i \leq j \\
	      -\infty, & \text{otherwise}
	    \end{cases}
\end{equation}
We use a chunk size of 32 ms and chunk shift of 16 ms, and verify that the computation time for a chunk is less that 16 ms on CPU.

\begin{table*}[t!]
\centering
\caption{STOI and PESQ comparisons between DP-SARNN and baseline models of a) complex spectral mapping, and b) time-domain enhancement.}
\begin{adjustbox}{width=0.72\textwidth}
\begin{tabular}{|c|c|c|c|cccc|cccc|cccc|cccc|}
\hline
\multirow{4}{*}{ \rotatebox{90}{Approach} } & \multirow{4}{*}{ \rotatebox{90}{Causal?} } & \multirow{4}{*}{ \rotatebox{90}{Real-time?} } & Metric & & \multicolumn{7}{c|}{ STOI (\%)} & \multicolumn{8}{c|}{ PESQ } \\
\cline{4-20}
& & & Test Noise & \multicolumn{4}{c|}{ Babble } & \multicolumn{4}{c|}{ Cafeteria } & \multicolumn{4}{c|}{ Babble } & \multicolumn{4}{c|}{ Cafeteria } \\
\cline{4-20}
& & & Test SNR & -5 db & 0 dB & 5 dB & AVG & -5 dB & 0 dB & 5 dB & AVG & -5 db & 0 dB & 5 dB & AVG & -5 dB & 0 dB & 5 dB & AVG \\
\cline{4-20}
& & & Mixture & 58.4 & 70.5 & 81.3 & 70.1 & 57.1 & 69.7 & 81.0 & 69.2 & 1.56 & 1.82 & 2.12 & 1.83 & 1.46 & 1.77 & 2.12 & 1.78 \\
\hline
\hline
\multirow{2}{*}{ a) } & \cmark & \cmark & GCRN \cite{tan2019learning} & 82.4 & 90.9 & 94.8 & 89.4 & 79.1 & 89.3 & 94.0 & 87.5 & 2.17 & 2.70 & 3.07 & 2.65 & 2.10 & 2.60 & 2.99 & 2.56 \\
& \xmark & \xmark & NC-GCRN \cite{tan2019learning} & 87.0 & 93.0 & 95.6 & 91.9 & 84.1 & 91.7 & 95.1 & 90.3 & 2.53 & 2.96 & 3.25 & 2.91 & 2.40 & 2.85 & 3.17 & 2.81 \\
\hline
\multirow{9}{*}{ b) } & \cmark & \xmark & SEGAN-T \cite{pascual2017segan} & 81.5 & 90.3 & 94.1 & 88.6 & 79.8 & 89.5 & 93.5 & 87.6 & 2.11 & 2.62 & 2.97 & 2.57 & 2.15 & 2.61 & 2.94 & 2.57 \\
& \cmark & \xmark& AECNN \cite{pandey2019new} & 82.6 & 91.5 & 95.1 & 89.7 & 81.1 & 90.7 & 94.5 & 88.8 & 2.21 & 2.80 & 3.17 & 2.73 & 2.23 & 2.76 & 3.12 & 2.70 \\
& \cmark & \cmark & TCNN \cite{pandey2019tcnn} & 82.8 & 91.3 & 94.8 & 89.6 & 80.6 & 89.8 & 94.0 & 88.1 & 2.18 & 2.70 & 3.06 & 2.65 & 2.14 & 2.62 & 2.98 & 2.58 \\
& \cmark & \cmark & DCN \cite{pandey2020dense} & 85.1 & 92.7 & 95.8 & 91.2 & 82.5 & 91.3 & 95.1 & 89.6 & 2.31 & 2.91 & 3.30 & 2.84 & 2.29 & 2.82 & 3.22 & 2.78 \\
& \cmark & \xmark & DP-LSTM & 87.5 & 93.7 & 96.2 & 92.5 & 83.8 & 92.1 & 95.5 & 90.4 & 2.50 & 3.04 & 3.36 & 2.97 & 2.35 & 2.91 & 3.29 & 2.85 \\
& \cmark & \cmark & DP-SALSTM & \textbf{88.5} & \textbf{94.3} & \textbf{96.6} & \textbf{93.1} & \textbf{84.7} & \textbf{92.7} & \textbf{95.8} & \textbf{91.1} & \textbf{2.54} & \textbf{3.10} & \textbf{3.42} & \textbf{3.02} & \textbf{2.41} & \textbf{2.94} & \textbf{3.32} & \textbf{2.89} \\
& \xmark & \xmark & NC-DCN & 89.0 & 94.3 & 96.6 & 93.3 & 85.6 & 93.0 & 95.9 & 91.5 & 2.71 & 3.18 & 3.48 & 3.12 & 2.56 & 3.07 & 3.39 & 3.01 \\
& \xmark & \xmark & DP-BLSTM & 90.08 & 94.7 & 96.7 & 93.8 & 86.8 & 93.5 & 96.1 & 92.1 & 2.82 & 3.23 & 3.48 & 3.18 & 2.65 & 3.13 & 3.43 & 3.07 \\
& \xmark & \xmark & DP-SABLSTM & \textbf{91.61} & \textbf{95.3} & \textbf{97.1} & \textbf{94.7} & \textbf{87.9} & \textbf{94.0} & \textbf{96.4} & \textbf{92.8} & \textbf{2.94} & \textbf{3.30} & \textbf{3.54} & \textbf{3.26} & \textbf{2.70} & \textbf{3.17} & \textbf{3.46} & \textbf{3.11} \\
\hline
\end{tabular}

\end{adjustbox}
\end{table*}

\begin{table}[b!]
\centering
\caption{Model size and processing time comparisons between DCN, DP-LSTM, and DP-SALSTM.}
\begin{adjustbox}{width=0.86\columnwidth}
\begin{tabular}{|c|c|c|c|}
\cline{2-4}
\multicolumn{1}{c|}{}&\ \ \ \ \ DCN \ \ \ \ \ & \ \ \ \ DP-LSTM\ \ \ \ & DP-SALSTM \\
\hline
(K, P) & - & (255, 127) & (63, 31) \\
(L, R) & (512, 256) & (4, 2) & (16, 8) \\
Chunk size (ms) & 32 & 32 & 32 \\
\# params (millions) & 5.6 & 3.37 & 6.49 \\
Avg time per chunk (ms) & 11.0 & 17.4 & 7.9 \\
\hline
Is real-time? & \cmark & \xmark & \cmark \\
\hline
\end{tabular}
\end{adjustbox}
\end{table}
\section{Experiments}
\subsection{Datasets}
We train speaker- and noise-independent models using a large number of noises and speakers. We use the WSJ0 SI-84 dataset \cite{paul1992design} consisting of 83 speakers (42 males and 41 females) in which 76 are used for training and the remaining 6 (3 males and 3 females) are used for evaluation.

320000 training mixtures are generated by adding random noise segments from a sound effect library of 10000 non-speech sounds (\url{www.sound-ideas.com}) at random SNRs from \{-5 dB, -4 dB, -3 dB, -2 dB, -1 dB, 0 dB\}. 

Test mixtures are generated by using two noises (babble and cafeteria) from an Auditec CD (\url{http://www.auditec.com}) at SNRs of -5 dB, 0 dB and 5 dB. We generate 150 mixtures for all pairs of test noises and test SNRs. For the validation set, we use 6 speakers from the training set (150 utterances) and mix it with factory noise at an SNR of -5 dB. 
\subsection{Baselines}
We compare DP-SARNN with a recently proposed gated convolutional recurrent network (GCRN) \cite{tan2019learning}, auto-encoder CNN (AECNN) \cite{pandey2019new}, speech enhancement generative adversarial network (SEGAN) \cite{pascual2017segan}, temporal convolutional neural network (TCNN) \cite{pandey2019tcnn}, and dual-path RNN (DP-RNN) \cite{luo2020dual}. DP-RNN is adopted for speech enhancement by using one decoder instead of two and replacing masking with mapping as in DP-SARNN. Non-causal DP-RNN and DP-SARNN use BLSTM RNN, whereas causal DP-RNN and DP-SARNN replace BLSTM with LSTM in inter-chunk RNN.  In our results (Table 1 and Table 2), causal DP-RNN and DP-SARNN are respectively denoted as DP-LSTM and DP-SALSTM, and non-causal DP-RNN and DP-SARNN are respectively denoted as DP-BLSTM and DP-SABLSTM.
\subsection{Experimental settings}
All the utterances are resampled to 16 kHz.  We use $L = 16$, $R = 8$, $K = 126$ for DP-SABLSTM, $K = 63$ for DP-SALSTM, $N=128$, and $H=256$. For DP-SABLSTM and DP-BLSTM, $K$ is chosen in a way so that $K \approx J$. Dropout rate in feedforward block is set to 5\%.  We use phase constrained magnitude (PCM) loss for training \cite{pandey2020dense} . All the models are trained for 15 epochs on 4 second long utterances with a batch size of 8. Mixed precision training \cite{micikevicius2018mixed} is utilized for faster training. Learning rate is set to 0.0002 for first 5 epochs and exponentially decayed afterwards at every epoch using a rate that results in a learning rate of 0.00002 in the last epoch. Gradient clipping is applied during training with a maximum $l^{2}$ norm of 3.

We observe short-time objective intelligibility (STOI) \cite{taal2011algorithm} score on the validation set after each epoch of training, and the model with maximum STOI is used for evaluation.
\subsection{Experimental results}
We compare all the models in terms of STOI whose values typically range between 0 and 1 and perceptual evaluation of speech quality (PESQ) whose values range from -0.5 to 4.5 \cite{rix2001perceptual}. 

First, we compare DP-SARNN with different baselines in Table 1. There are four real-time models in Table 1: GCRN, TCNN, DCN, and DP-SALSTM. The performance trend for these models is GCRN $<$ TCNN $<$ DCN $<$ DP-SALSTM. In particular, on average, DP-SALSTM improves scores over DCN by 1.7\% in STOI and 0.15 in PESQ, where best improvements are obtained for the difficult SNR of -5 dB.

A similar behavior is observed for non real-time models, in which case the performance trend is SEGAN-T $<$ AECNN $<$ NC-GCRN $<$ NC-DCN $<$ DP-LSTM $<$ DP-BLSTM $< $DP-SABLSTM. Note that SEGAN, AECNN and DP-LSTM are causal systems but not real-time because AECNN and SEGAN respectively use frame size of 128 ms and 1024 ms, and DP-LSTM does not satisfy the latency constraint (Table 2).

Next, we compare top three causal systems, DP-SALSTM, DP-LSTM, and DCN in terms of number of parameters and average computation time for a signal chunk of 32 ms on a 2.4 GHz quad core machine with Intel Xeon E5-2680 v4 processors and 4GB RAM. The results are given in Table 2. The computation time is in order DP-SALSTM $<$ DCN $<$ DP-LSTM. Even though DP-LSTM is causal, and has fewer parameters, it takes 17 ms CPU time to process a signal chunk of 32 ms, which is greater than the chunk shift (16 ms), and hence it is a non-real-time system. DP-SALSTM is faster than DP-LSTM because it uses a frame shift of 8 instead of 2, and as a result, it needs to process four times less number of frames. DP-LSTM can be converted to a real-time model by increasing the frame size and frame shift, but it will lead to performance degradation as in \cite{luo2020dual}. Number of parameters in different models are in order  DP-LSTM $<$ DCN $<$ DP-SALSTM.

We plan to explore DP-RNN and DP-SARNN for cross-corpus generalization by training on LibriSpeech \cite{panayotov2015librispeech} as in \cite{pandey2020cross, pandey2020learning}. 
\vspace{-11pt}
\section{Conclusions}
We have proposed a novel dual-path self-attention RNN for time-domain speech enhancement. The proposed DP-SARNN augments RNNs in DP-RNN with attention. Adding attention in DP-RNN leads to improved enhancement with a four times frame shift, resulting in a low-latency model. As a result, we have developed a real-time version of DP-SARNN that is not only also faster than but also outperforms existing approaches. Future work includes exploring and improving DP-SARNN for cross-corpus generalization.

\bibliographystyle{IEEEbib}
\bibliography{mybib}

\end{document}